\documentclass[prb,aps,twocolumn,showpacs]{revtex4}
 \usepackage{graphicx}

\begin{document}
 
 \title{Wigner Crystallization in inhomogeneous one dimensional wires}
 \author{Erich J. Mueller}
\affiliation{Laboratory of Atomic and Solid State Physics, Cornell University, Ithaca, New York 14853}
\date{\today}

\begin{abstract}
We explore the theory of electrons confined by one dimensional power law potentials.
We calculate the density profile 
in the high density electron gas,  the low density Wigner crystal, and the intermediate regime.  We extract the momentum space wavefunction of the electron at the Fermi surface, which can be measured in experiments on tunneling between parallel wires.   The onset of localization leads to a dramatic broadening of the momentum space wavefunction together with pronounced sharpening (in energy) of the tunneling spectrum.
\end{abstract}
 \maketitle

Advances in microprocessing have enabled scientists to construct ultra-high mobility one dimensional wires with transverse size $d_\perp\approx20$nm, which is sufficiently small to freeze out all transverse electronic motion \cite{wires} . By allowing tunneling between two such parallel wires, Auslaender et al. \cite{aus1}  have been able to map out the spectrum of elementary excitations, providing good evidence of spin-charge separation.  Further experiments have found evidence of electronic localization in these wires when a gate electrode is used to deplete the density of electrons \cite{yacobytalk,auslaendernature}.  
This localization is possibly caused by an interaction driven transition into a Wigner crystal state \cite{wigner,oned}, where each electron is localized by repulsion from its neighbors. Here we investigate the theory of such crystallization in one dimensional wires.

The wires used in these experiments are not infinitely long.  Their finite length, which is controlled by electrostatic potentials, leads to striking fringes in the tunneling spectrum \cite{tserkovnyak}.  These fringes cannot be explained by particles bounded by `infinite walls', but require a more accurate modeling of the confinement. Tserkovnyak et al. \cite{tserkovnyak} found that  the experimental results were consistent with a power-law potential $V_{\rm ext}(x)=V_0 x^\beta,$ with $6<\beta<7$.  In discussing experimental signatures of crystallization in these systems, we must take into account this same physics, and explore the interplay between the external confining potential and the interparticle Coulomb repulsion.  Due to the influence of the gate electrode, the electrons in the localization experiment\cite{yacobytalk,auslaendernature} feel a complicated potential, with two minima separated by a barrier.  For most of our analysis we use the simpler power law potential.

We are interested in gaining an intuitive understanding the crossover between electron liquids and solids in these inhomogeneous wires.  We therefore choose to work with the simplest techniques which can capture the qualitative features of these experiments.  We use
a Thomas-Fermi description of the high density electron gas (sec. \ref{sec:tf}); a semiclassical description of the Wigner crystal (sec. \ref{classical}); and a Hartree-Fock description of the cross-over (sec. \ref{hf}).   These models provide direct insight into the experimental signatures of this transition.  They do not, however, give direct information about excitations, transport properties, or fractionalization of the electrons into spin and charge degrees of freedom. 

Our approach complements prior theoretical work, which predominantly treated homogeneous infinite systems \cite{infinite}, small numbers of electrons \cite{small}, or infinite square well potentials \cite{square}.  Particular note should be taken of the recent work by Fiete et al.\cite{halperin} which analyzes the experiments discussed here\cite{yacobytalk,auslaendernature} in the limits of extremely small and extremely large particle numbers.

Our main accomplishment is that we are able to follow the evolution of the quantum state from the weakly interacting high density regime to the strongly interacting low density regime.  We describe the electronic density profile, and see how Friedel oscillations evolve into Wigner-crystal correlations.  In all of these regimes, we predict the tunnelling spectrum between parallel wires, giving qualitative explanation of the striking features in recent experiments\cite{yacobytalk,auslaendernature}.

%



\section{Model}\label{model}
We consider a gas of electrons with $\sigma=2$ spin states confined to a one dimensional wire, experiencing an external potential $V_{\rm ext}(x)$, where $x$ is the coordinate along the wire.  The electrons interact through pairwise three dimensional coulomb interactions, $U_{3d}(r_i-r_j)=e^2/(4\pi \epsilon|r_i-r_j|)=(\hbar^2/ m^* a)1/ |r_i-r_j|$, where $\epsilon$ is the dielectric constant of the medium and $m^*$ is the electrons effective mass.  Using parameters for n-doped GaAs ($\epsilon\approx13,m^*\approx0.067 m_e$, where $m_e$ is the free-space electron mass), the effective Bohr radius is quite large, $a\approx10^{-8}$m.  In this wire, the electronic wavefunction has a transverse size $d_\perp\approx 20$nm.  Integrating out these transverse dimensions, the electrons feel a regularized Coulomb interaction $U(r)=(\hbar^2/m^*a) f_{d_\perp}(r_i-r_j)$.  Two useful regularizations are $f_d^{(1)}(x)=(x^2+d^2)^{-1/2}$ and
$f_d^{(2)}(x)={\rm Min}(|x|^{-1},d^{-1})$, where ${\rm Min(x,y)}$ is the smaller of $x$ and $y$.
As $d_\perp\to0$ the exact form of the regularization becomes unimportant.  Here we mostly rely on $f^{(2)}$.

One estimates the importance of correlations caused by the Coulomb interactions by comparing the interaction energy between two neighboring particles $E_{\rm int}= \hbar^2n/ a m^*$
to the Fermi (kinetic) energy
${\cal E}_f = k_f^2/2m^* = 2\pi^2 \hbar^2 n^2/\sigma m^*$, where $n$ is the one dimensional density of particles, and we have assumed $nd_\perp\ll1$.  Kinetic energy dominates at high densities, when $n a\gg1/2\pi^2$.  In that limit one expects to find a liquid state, where electrons are delocalized.  Conversely, when $na \ll1/2\pi^2$, the Coulomb interaction dominates, and a Wigner crystal should be formed.

As a compromise between
experimental relevance and simplicity we concentrate on power law
potentials $V_{\rm ext}(x)=(\hbar^2/2 m^*) x^\beta w^{-\beta-2}$, where $w$, which parameterizes the potential strength,  roughly coincides with the size of the single particle ground state in this potential.  Figure~\ref{fig:tunnel}(b) illustrates the structure of this potential for different values of $\beta$.  More complicated potentials are briefly discussed in appendix~\ref{doublewell}.

For the relatively flat (large $\beta$) potentials used in experiments, $w$ also roughly coincides with the   
`lithographic length,' $L_{\rm lith}$, which is the physical distance between the gates which generate the potential barriers.  The limit $\beta\to\infty$ corresponds to an infinite square well.
Analysis of experimental data by Tserkovnyak et al. \cite{tserkovnyak} show that  $w\approx3\mu m$.  Taking the density of electrons to be $n\approx N/w$, one expects crystallization when $N<N^*$, with $N^*\sim 20$.  

\section{Tunneling between parallel wires}\label{tunnel}
The experiments of Auslaender et al.\cite{aus1} measure the tunneling current between the short wire described in section~\ref{model}, and a parallel long wire, which we will take to be infinite and uniform.  
Momentum is conserved in the tunneling.  

We will make the following approximations to analyze this experiment.  (1) We will treat the tunneling within perturbation theory.  This is an important limit to study, though,
as will be shown below, the experiments are in a regime where higher order processes may be playing a role.  (2) We will neglect correlations between electrons in the two wires.  One expects that this approximation will break down in the extremely low density limt, when the separation between electrons in one of the wires becomes comparable to the interwire spacing.  (3) In the bulk of this paper we use an independent electron picture to study the electrons in the short wire.  The consequences and validity of this approximation are discussed in section~\ref{indep} and are explored in more detail in appendix~\ref{two}.  We expect that in the limit where the splitting between different spin states is small compared to the temperature or linewidths that 
the independent electron picture correctly predicts the magnetic field dependence of the tunneling current.  However, this approximation results in a vast overestimation of the overall amplitude of the current.

\begin{figure}[tbph]
\includegraphics[width=\columnwidth]{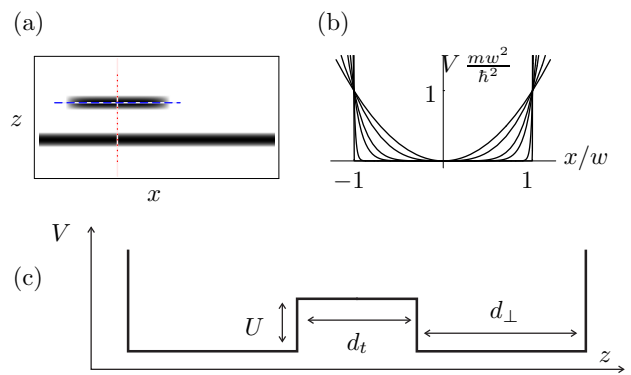}
\caption{(Color Online) Model of potential. (a) Two dimensional grayscale plot, where darker colors represent lower potential energy. (b) Potential along one wire [dashed blue line in (a)] for $\beta=2,3,6,10,50,1000$. (c) Potential transverse to the two wires [dotted red line in (b)].}\label{fig:tunnel}
\end{figure}

We consider the geometry sketched in figure~\ref{fig:tunnel}. 
The wires are separated by $d_t\approx6$nm, and the barrier separating the wires has height $U\approx300$meV, arising from the conduction band offset between the GaAs wires, and the intervening AlGaAs.  Using gate electrodes, the experimentalists control the chemical potential difference $\delta V$ between the wires.\cite{mufoot}  A magnetic field, perpendicular to the wires, gives a momentum kick $Q=e B d_t/\hbar$ to a tunneling electron\cite{aus1}.
We therefore consider the tunneling Hamiltonian,
\begin{equation}
\hat H_t=-T\sum_\sigma\int\frac{dk}{2\pi} e^{i (\delta V) t}\hat\phi^\dagger_\sigma({k+Q}) \hat\psi_\sigma(k) + {\rm H. C.},
\end{equation}
where $\hat \psi_\sigma(k)=\int dx\,e^{-ikx}\,\hat\psi_\sigma(x)$ and $\hat \phi_\sigma(k)=\int dx\,e^{-ik x}\,\hat\phi_\sigma(x)$ are respectively the operators which annihilate particles with momentum $k$ and spin $\sigma$ in the short and long wire.
The magnitude of the tunneling matrix element $T$ is estimated by examining the energy states in the double square well geometry sketched in figure~\ref{fig:tunnel}.  This one-dimensional potential is a crude model of the physics transverse to the wires.  We can identify $T\approx \delta E/2$, where $\delta E$ is the energy splitting between the two lowest energy single particle states.  
In the limit of a deep, wide barrier ($U\gg (\hbar^2/m^*d_t^2),(\hbar^2/m^*d_\perp^2)$), one finds 
\begin{equation}
\delta E/2 = \frac{2 \pi^2\hbar^2}{m^* d_\perp^3\kappa}  e^{-\kappa d_t},
\end{equation}
where $\hbar^2\kappa^2/2m^*=U$ (so numerically $\kappa\approx 0.2/{\rm nm}$).
which gives $T\approx30$meV, which should be compared to the spacing of states transverse to the wire, $\delta E =4\pi^2\hbar^2/m^* d_\perp^2\approx 100$meV. 
%
 Due to this separation of scales,  we treat $H_t$ perturbatively.  Note, however, that the ratio $T/\delta E$ is not extremely small, so that quantitative comparison with experiment would require going beyond lowest order in perturbation theory.

\subsection{Formal expression for tunneling current}
The current operator, defined by
$ I(t)=-{\partial N_\psi}/{\partial_t}$, where $\hat N_\psi=\int\frac{dk}{2\pi}\sum_\sigma \hat\psi_\sigma^\dagger(k)\hat\psi_\sigma(k)$ is the number of particles in the short wire, is given by
\begin{equation}
\hat I(t)=-T\int\frac{dk}{2\pi}\sum_{\sigma}{\rm Im} \left[
e^{i\delta V t}\hat\phi_\sigma^\dagger(k+Q,t)\hat\psi_\sigma(k,t)
\right].
\end{equation}
 To lowest order in $T$, the 
 tunneling current is
\begin{eqnarray}
\langle I(t)\rangle &=&\int\!d\tau\frac{1}{i}\theta(t-\tau)\langle [ I(t),H_t(\tau)]\rangle\\
&=& T^2 A_I(Q,\omega=\delta V),
\end{eqnarray}
where the spectral density of current fluctuations is related to the retarded current response function by $A_I(q,\omega)\equiv2{\rm Im} \chi_I^R(q,\omega),$ where
$\chi_I^R(q,\omega)=\int\! dt\,e^{i\omega t}\,(\theta(t)/{i})\,[\chi_I^>(q,t)-\chi_I^<(q,t)],$ with
$\chi_I^>(q,t)=\int\frac{dk}{2\pi}\frac{dk^\prime}{2\pi}\sum_{\sigma}
 \langle \hat\phi^\dagger_\sigma(k+q,t)\hat\psi_\sigma(k,t)
\hat \psi_\sigma^\dagger(k^\prime,0)\hat\phi_\sigma(k^\prime+q,0)
 \rangle$ and
$\chi_I^<(q,t) =\int\frac{dk}{2\pi}\frac{k^\prime}{2\pi}\sum_{\sigma}
 \langle  \hat\psi_\sigma^\dagger(k^\prime,0)\hat\phi_\sigma(k^\prime+q,0)
 \hat\phi^\dagger_\sigma(k+q,t)\hat\psi_\sigma(k,t)
\rangle.$
To simplify this expression we assume that in the electrons in the two wires are independent. 
Thus we neglect correlations between the two wires caused by the Coulomb interaction.  Such correlations are likely to be present in the extreme Wigner crystal limit, where the spacing between electrons in one of the wires is large compared to the wire separation.  Quantitative comparison with experiments in this limit would require including these correlations, which will  partially screen the long-range part of the Coulomb interaction.

Within our approximation
the expectation values can be factored, leading to
\begin{widetext}
\begin{eqnarray}\label{afac}
A_I(q,\omega)&=&\int\frac{dk}{2\pi}\frac{dk^\prime}{2\pi}\sum_{\sigma}\int\frac{d\nu}{2\pi} 
\left[G_{\phi\sigma}^<(k,k^\prime;\nu)G_{\psi\sigma}^>(k+q,k^\prime+q;\nu-\omega)\right.
\left.-
G_{\phi\sigma}^>(k,k^\prime;\nu)G_{\psi\sigma}^<(k+q,k^\prime+q;\nu-\omega)
\right]\\
&=&\int\frac{dk}{2\pi}\frac{dk^\prime}{2\pi}\sum_{\sigma}\int\frac{d\nu}{2\pi} 
\left[f(\nu)-f(\nu-\omega)\right]
A_{\psi\sigma}(k,k^\prime;\nu) A_{\phi\sigma}(k+q,k^\prime+q;\nu-\omega),
\end{eqnarray}
\end{widetext}
where the single particle Greens functions $G^{>/<}_{\phi\,\sigma}(k,k^\prime;\omega)=\int dt\,e^{i\omega t} G^{>/<}_{\phi\,\sigma}(k,k^\prime;t),$ are given by $G^>_{\phi\,\sigma}(k,k^\prime;t)=\langle \phi_\sigma(k,t)\phi_\sigma^\dagger(k^\prime,0)\rangle,  G^<_{\phi\,\sigma}(k,k^\prime;t)=-\langle \phi_\sigma^\dagger(k^\prime,0)\phi_\sigma(k,t)\rangle,$ and equivalent expressions hold for $G^{>/<}_{\psi\,\sigma}(k,k^\prime,\omega)$.  These Greens functions are related to the appropriate single particle spectral density by $G^>(\omega)=[1-f(\omega)]A(\omega)$ and $G^<(\omega)=f(\omega)A(\omega)$, where $f(\omega)=[e^{\beta(\omega-\mu)}+1]^{-1}$ is the Fermi function.

If we use the free electron spectral density for the long wire, $A_{\phi\,\sigma}(k,k^\prime;\omega)=(2\pi)^2 \delta(k-k^\prime) \delta(\omega-k^2/2m^*),$ one arrives at the simple result that the tunneling current is a direct measure of the single particle spectral density in the short wire,
\begin{eqnarray}
A_I(\omega,q)&=&
\int\frac{dk}{2\pi}\sum_{\sigma}\left[f(\omega+k^2/2m^*)-f(k^2/2m^*)\right] \nonumber\\&&\qquad\times A_{\psi\,\sigma}(k-q,\omega+k^2/2m^*)\label{specdens}
\\
&\approx&\nonumber
\beta\omega \int\frac{dk}{2\pi}\sum_{\sigma} f_k(1-f_k)A_{\psi\,\sigma}(k-q,k^2/2m^*),
\end{eqnarray}
where the last line neglects terms of order $\omega^2$,  and uses $f_k=f(k^2/2m^*)$.

At zero temperature, the Fermi functions become step functions and to lowest order in $\omega$, equation (\ref{specdens}) becomes
\begin{eqnarray}\nonumber
A_I(\omega,q)&=&
\frac{\omega}{2\pi}\sqrt{\frac{m^*}{2{\cal E}_f}} \sum_\sigma \left[
A_{\psi\,\sigma}(\sqrt{2m^*{\cal E}_f}-q,{\cal E}_f)\right.\\&&\quad\left.+A_{\psi\,\sigma}(-\sqrt{2m^*{\cal E}_f}-q,{\cal E}_f)
\right].
\end{eqnarray}

\subsection{General Features}\label{orthog}
From the definition $A_{\psi\,\sigma}(k,\omega)=G^>_{\psi\,\sigma}(k,\omega)+G^<_{\psi,\sigma}(k,\omega)$, the single particle spectral density can be written as
\begin{widetext}
\begin{eqnarray}\label{sd}
A_{\psi\,\sigma}(k,\omega)&=&
\frac{2\pi}{Z}\sum_{if} e^{-\beta(E_i-\mu N_i)} \left[
\left|\langle i | \hat \psi_\sigma(k)|f\rangle\right|^2 \delta(\omega-({ E}_f-E_i))-
\left|\langle f | \hat \psi_\sigma(k)|i\rangle\right|^2 \delta(\omega-(E_i-{ E}_f))
\right]
\end{eqnarray}
\end{widetext}
where $|i\rangle$ represents a many body state containing $N_i$ electrons and possessing  energy $E_i$.  Normalization is given by the grand partition function  $Z=\sum_i e^{-\beta (E_i-\mu N_i)}$.
At zero temperature, the sum over $i$ is omitted, and $|i\rangle$ is replace by the ground state.  

One thus sees that current flows whenever momentum and energy can be conserved in the tunneling process.  The magnitude of the current is set by the degree of overlap between states with different particle number.  

When the electrons become localized in a Wigner crystal, the overlap between states with different particle number becomes extremely small (because removing a particle results in rearrangement of the entire crystal -- a form of orthogonality catastrophe).  Consequently, the average tunneling current should drop.  This drop in average current is accompanied by a sharpening of line-widths and does not lead to a complete loss of the experimental signal.  Instead, it leads to ``coulomb-blockade" type peaks, where current only flows for discrete values of $\delta V$.

One will see discrete peaks if the line width is small compared to the level spacing in the short wire, $\delta E\sim (2\pi)^2 \hbar^2/m^* w^2\sim0.05$meV.  Assuming that the lifetime is only due to coupling between the two wires, then the line width is given by an expression very similar to that for the tunneling current.  Since $T$ is large on the scale of $\delta E$, one will only see these discrete peaks once the overlaps in (\ref{sd}) become extremely small.

\subsection{Independent particle approximation}\label{indep}
In an independent particle picture (such as Hartree-Fock), $A_{\psi\,\sigma}(k,\omega)=\sum_j |\phi_{j\,\sigma}(k)|^2\,2\pi\delta(\omega-E_{j\,\sigma})$, where $\phi_{j,\sigma}(k)$ is the momentum space wavefunction of the $j$'s single electron orbital of spin $\sigma$ particles, with energy $E_{j\,\sigma}$.  In general $\phi_{j\,\uparrow}\neq\phi_{j\,\downarrow}$, and the orbitals are chosen self-consistently.
Within such an approximation the tunneling current is
\begin{equation}\label{indi}
I=I_0 
\sum_{n\sigma} \left| \phi_{n\,\sigma}(k
)\right|^2 \delta(E_{n\,\sigma}-{\cal E}_f),
\end{equation}
where, $I_0=T^2{\delta V}/({2\pi})\sqrt{ m^*/2 {\cal E}_f}$, $k=\sqrt{2m^* {\cal E}_f}-Q$, and 
we have assumed that $Q=e B d_t/\hbar>0$.  The tunneling current therefore measures the momentum space density of the particles at the Fermi surface.  If there are no particles at the Fermi surface, then no current flows.

The delta functions in (\ref{indi}) are broadened by finite lifetime and finite temperature.  In particular if there are two different states $i,j$ with energies $E_{i/j}$, and inverse lifetimes $\Gamma$, which satisfy $|E_{i/j}-{\cal E}_f|\ll\Gamma$ or $|E_{i/j}-{\cal E}_f|\ll k_B T$, then the current is proportional to the incoherent sum
\begin{equation}
I\propto |\phi_i(k)|^2+|\phi_j(k)|^2.
\end{equation}
This feature will be important in high symmetry situations where there is a near degeneracy (see, for example, section~\ref{fringe}).

In the remaining sections of this paper, we calculate $\phi_j(k)$.

In its conventional form, the independent particle approximation does not include the effects of orthogonality between many-body states of different particle number, as described in section~\ref{orthog}.  In appendix~\ref{two} we explore a simple example which lets us systematically study these effects.  We find that the omitted physics predominately modifies the overall scale of the tunneling current, and can be partially accounted for by renormalizing $I_0$ in equation (\ref{indi}).  Hartree-Fock also misses fine-detail fringe structure in the magnetic field dependance of the tunneling current.  The prediction of the envelope of the tunneling current appears to be qualitatively correct.



%


\section{High density: Thomas-Fermi}\label{sec:tf}
\begin{figure}
\includegraphics[width=\columnwidth]{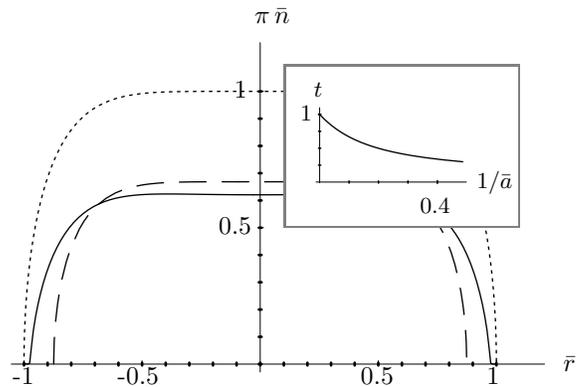}
\caption{Interaction induced change of dimensionless density $\bar n$ of spin-1/2 electron gas confined to a one-dimensional potential $V(x)\propto x^6$.  Interaction parameters are $\bar a=10$ and $\bar d=10^{-3}$. 
Solid line: Equation (\ref{tf2}) using regularization $f^{(2)}$; dashed line: approximation described above eq.~(\ref{trans}); dotted line: noninteracting profile.  
Inset: Renormalized dimensionless cloud radius $t$ as a function of  $1/\bar a$ from solving eq.~(\ref{trans}). }\label{showt}
\end{figure}
The simplest picture of a gas of electrons in a confined geometry comes from the Thomas-Fermi (local density) approximation, where the system is described by a local chemical potential $\mu(x)=\mu_0-V_{\rm eff}(x)$.  The global chemical potential is $\mu_0=\hbar^2k_0^2/2m^*$, and the effective potential $V_{\rm eff}(x)$ includes both the external field and the interactions between the particles.  Within the Hartree approximation, which is valid at high densities $ n a \gg 1$, the effective potential is
\begin{equation}
V_{\rm eff}(x)=\int \!dx^\prime\,U(x-x^\prime) n(x^\prime).
\end{equation}
The chemical potential is then related to the density by the relationship for a homogeneous gas, $\mu(r)=2\pi^2 \hbar^2 n^2/\sigma^2 m^*$.  Self-consistency requires that the
density obeys a nonlinear integral equation,
\begin{equation}\label{tf}
\mu_0 = V_{\rm ext}(x) + \int\!dx^\prime U(x-x^\prime) n(r^\prime) + \frac{2\pi^2\hbar^2}{\sigma^2m^*} n^2(r).
\end{equation}
This local density approximation only makes sense if the density changes slowly compared to the interparticle spacing [$(\partial_x n(x))/n(x)^2\ll1$].  
We write (\ref{tf}) in dimensionless form by introducing $k_0^2=2 m^*\mu_0/\hbar^2$,
$R^\beta=k_0^2 w^{2+\beta}$, $\bar r=r/R$, $\bar a=k_0 a$, $\bar n= k_0 n$, $\bar d= d/R$,
\begin{equation}\label{tf2}
\frac{4\pi^2}{\sigma^2}{\bar n^2(r)}=
1-\bar r^\beta - \frac{2}{\bar a}\int_{-1}^1 f_{\bar d}(\bar r-\bar r^\prime){ n(\bar r^\prime) d\bar r^\prime}.
\end{equation}
At fixed chemical potential the interactions can only reduce the density.  Therefore the density is always bounded by the noninteracting result, $\bar n_0=(\sigma/2\pi)\sqrt{1-\bar r^\beta}$.  In particular, the density always vanishes for $\bar r>1$.  For very steep potentials, $\beta\to\infty$, the noninteracting Thomas-Fermi radius $R$ approaches  the trap length,  $R/w\approx 1+2\beta^{-1}\log(k_0 w)+{\cal O}(\beta^{-2})$.



For $a\gg1$, equation (\ref{tf2}) can be solved iteratively \cite{it}.  In this limit, the interactions predominately renormalize the chemical potential.  Approximating the integral in (\ref{tf2}) by its value at $\bar r=0$, one self-consistently finds that  $(2\pi/\sigma) \bar n(\bar r)\approx \sqrt{t^2-\bar r^\beta}$ where $t$ solves the transcendental equation
\begin{equation}\label{trans}
t^2=1-\frac{\sigma}{\pi \bar a} t(A+4\log t).
\end{equation}
The parameter $A$ depends logarithmically on the cut-off $d_\perp$.  For the regularization $f_2$, 
$A=1-{4}/{\beta}+({2}/{\beta})\log({4}/{d_\perp^\beta})$.  Figure~\ref{showt} compares this approximation to the numerical solution of (\ref{tf2}), and the inset shows $t$ as a function of $\bar a$.


\subsection{Wavefunction of last occupied state}
\begin{figure}
\includegraphics[width=\columnwidth]{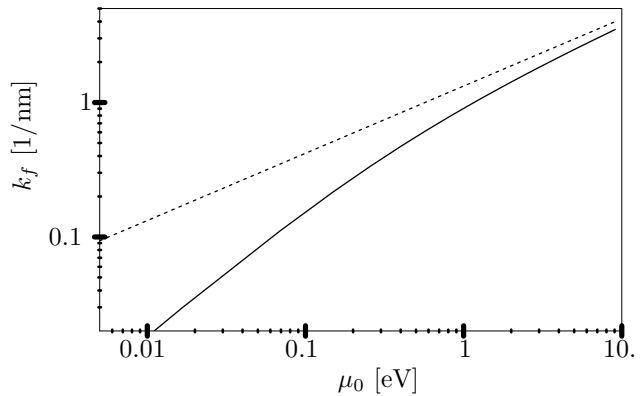}
\caption{
Peak wavevector $k_f=tk_0$ of wavefunction at fermi surface as a function of chemical potential $\mu_0=\hbar^2k_0^2/2m^*$ within the Thomas-Fermi approximation.  Dotted line shows noninteracting result,  $t=1$.  Solid line includes Coulomb interactions with $a=10 nm$, $d_\perp=20 nm$, in a power law trap with $\beta=6$ and $w=3\mu$m.  This approximation will break down when $k_f\lesssim1/a$.
}\label{kplot}
\end{figure}

As previously explained, within an independent particle picture
the
tunneling current between two parallel wires
 is proportional to the momentum space wavefunction of the highest occupied single particle state.  This momentum space wavefunction will be strongly peaked about $k_f(r=0)$, which can be approximated as $k_0 t$, where $\hbar^2 k_0^2/2m^*=\mu_0$, and $t$ is given by eq.~(\ref{trans}).  Figure~\ref{kplot} shows $k_f\equiv k_f(r=0)$ as a function of $\mu_0$ for reasonable parameters.

To go beyond this single wavevector approximation, we use semiclassical means to calculate the wavefunction of the state at the Fermi surface.  We find the modulus  by imagining that we reduce the chemical potential by a very small amount, so that this last state goes from filled to empty.  The density change must coincide with the density $n_f(r)$ of that last single particle state, implying that
\begin{equation}\label{dens}
n_f(x) = |\psi_f(x)|^2=\Omega \frac{\partial n(r)}{\partial\mu_0},
\end{equation}
where $\Omega$ is a normalization constant determined by setting $\int n_f(x)\,dx=1$.
This wavefunction should have a phase whose derivative gives the local fermi momentum.
\begin{equation}\label{phase}
\frac{\partial\varphi_f(x)}{\partial x} = \pm\sqrt{2m^*\mu(x)}\equiv k_f(x).
\end{equation}
The $+$ and $-$ solutions correspond to left-moving and right moving waves.  In principle, one must combine these two solutions to form a standing wave.
Within the Hartree approximation, this procedure is equivalent to
 finding the WKB wavefunction in the self-consistent potential.  We do not discuss the details as 
 when we compute the momentum space wavefunction, only the structure near $k=0$ will be affected by how we superimpose these two solutions.

In the case where the Coulomb interaction can be neglected,  equation (\ref{dens}) reduces to
$|\psi_f(x)|\propto k_f(x)^{-1/2}\propto1/\sqrt{ \mu_0-V_{\rm ext}(x)}$, and equation (\ref{phase}) becomes $\partial\varphi/\partial x = k_f(x)$.  Not surprisingly, in this limit $\psi_f(x)$ is {\em exactly} the WKB wavefunction for an electron with energy $\mu_0$ in the potential $V_{\rm ext}(x)$.  Our arguments therefore reduce to
the semiclassical arguments of  Tserkovnyak et al. \cite{tserkovnyak} for the noninteracting gas.
The analysis is very similar for our approximation where $\bar n^2\propto t^2-\bar r^\beta$.  Since interactions only renormalize the chemical potential, the derivative $\partial n^2(r)/\partial\mu_0$ is a constant, independent of space, and we can use the noninteracting result with a renormalized chemical potential.  That is, we take $k_f(x)=\sqrt{t^2 \mu_0-V_{\rm ext}(x)}=k_0\sqrt{t^2-\bar x^\beta}$.

Following Tserkovnyak et al. \cite{tserkovnyak}, we calculate the momentum space wavefunction, $\phi_f(q)=\int\! dx\, \psi_f(x) e^{-i q x}$ by  Laplace's method.  
The Fourier integral is dominated by regions of space near where $q=\pm k_f(x)$.  
Moreover, since $\phi_f(q)$ is peaked around $q=\pm k_f(0)$, we can
 expand $\varphi(x)$ near $x=0$.  Since $|\phi_f(q)|$ is symmetric, 
 there is no loss of generality in taking $q>0$.
We introduce $p=(k_f(0)-q) X$, with $X^{\beta+1}=2k_f w^{\beta+2}$.
Changing variables to $y=p x/X$, and assuming $(k_f(0)-q) \ll k_f(0)$, we find
\begin{equation}\label{airy}
\frac{\phi_f(q)}{\Omega^\prime }={\rm Re}\left[
\int_0^\infty\!\!\!\!dy
\exp\left(i p y - i \frac{y^{\beta+1}}{\beta+1}\right)
\right],
\end{equation}
where 
$\Omega^\prime$ is a normalization constant.  In the case $\beta=2$,  the integral in (\ref{airy}) is an Airy function, though for other values of $\beta$ it is not a familiar special function.  The argument of the exponential, $\chi$, is stationary at $y=y_0=p^{1/\beta}$.  The curvature is $\chi''(y_0)=-i \beta p^{1-1/\beta}$.  This is an isolated saddle point when $y_0^2|\chi''(y_0)|=\beta |p|^{1+1/\beta}\gg1,$ in which case 
\begin{equation}\label{ai2}
\frac{\phi_f(q)}{\Omega^\prime} =
{\rm Re}\left[
\sqrt{\frac{2\pi}{i \beta p^{1-1/\beta}}}\exp\left[i\frac{\beta p^{1+1/\beta}}{1+\beta}\right]
\right] .
\end{equation}
By deforming the contour of integration, this equation holds for both $p>0$ and $p<0$ (in the latter case one should take the principle branch, giving $p$ a small negative imaginary part).  From (\ref{ai2}), one sees that if $\beta\neq\infty$, 
the wave function falls off exponentially as $p\to-\infty$.
For $p\ll \beta^{\beta/(1+\beta)}$, instead of expanding about the saddle point, we expand about $y=0$, finding as $p\to0$,
\begin{eqnarray}\label{sp}
\frac{\phi_f(q)}{\Omega^\prime}&=&{\rm Re}\left[
(\beta+1)^{-\beta/(\beta+1)}
\Gamma\left(\frac{1}{\beta+1}\right)
e^\Theta\right]\\\nonumber
\Theta&=&
 p\,(\beta+1)^{1/(\beta+1)} 
\frac{\Gamma\left(\frac{2}{\beta+1}\right)}
{\Gamma\left(\frac{1}{\beta+1}\right)}
i^{1/(\beta+1)}\\\nonumber&&-i(\pi/2)(1/(\beta+1)).
\end{eqnarray}



\section{Low Density: Wigner Crystal}\label{classical}
In section~\ref{sec:tf}, we neglected the fact that electrons are discrete entities and that an electron does not interact with itself.  Including this discreteness, the interaction energy can be reduced by localizing each electron to a small volume -- maximizing the distance between the charge distribution of neighboring particles.  Since such localization costs kinetic energy, it is only favorable in the low density gas where interaction energy dominates over kinetic energy.

To discuss this phase, we imagine that we have $N$ electrons, localized at positions $x_j$ where $j=1,2,\ldots,N$.  Neglecting kinetic terms, the energy of a given configuration is 
\begin{equation}\label{wen}
E=\sum_j V_{\rm ext}(x_j)+(1/2)\sum_{i\neq j} U(x_i-x_j).  
\end{equation}
As with the profile in the dense limit (fig.~\ref{showt}), we numerically find that
the resulting density is quite flat, and is well approximated by evenly spaced electrons, $x_j n\approx j-(N+1)/2$, with uniform density $n$.  Minimizing (\ref{wen}) with respect to $n$ yields
\begin{eqnarray}
n^{-1}&=&\frac{2 w}{N} \left[
\frac{\beta+1}{2 \beta} \frac{w}{a} N^2\left(
\log N-\frac{1}{2}
\right)
\right]^{1/(\beta+1)}\\\nonumber
&\approx&
\frac{2 w}{N} \left[
1+\frac{1}{\beta+1}\log\left(
\frac{w N^2\log N}{2 a}
\right)+{\cal O}(\beta^{-2})
\right]
\end{eqnarray}


In this Wigner crystal state, each electron is confined to a length $\ell<n^{-1}$.  To calculate this length, and the wavefunction of the localized electron, we expand the potential in eq.~(\ref{wen}) for small fluctuations in the position of one of the particles,
\begin{equation}
\frac{\partial^2 V}{\partial x_j^2}= V_{\rm ext}''(x_j)+\frac{2 \hbar^2}{m^* a}\sum_{i\neq j} |x_i-x_j|^{-3}.
\end{equation} 
We approximate the sum over a finite number of particles by that of an infinite chain,
taking $\sum_{i\neq j} |x_i-x_j|^{-3}\approx 2 n^3 \sum_{l=1}^\infty l^{-3}=2\zeta(3)n^3$.  Numerically, the Riemann zeta function is $\zeta(3)\approx1.2$.  The curvature of the external potential is generically negligible in this limit, hence each electron is trapped by a harmonic oscillator potential with $m^*\omega^2_{\rm wc}=(\hbar^2/m^*a)4\zeta(3)n^3$ and has a wavefunction
 $\psi_j(x)\propto \exp(-(x-x_j)^2/2\ell^2)$ with $\ell^4=(\hbar/m^*\omega_{\rm wc})^2=a n^{-3}/(4\zeta(3))$.  This implies that the momentum space wavefunction has modulus
$|\psi_j(k)|^2\propto \exp(-k^2\ell^2)$.

Comparing this momentum space wavefunction to the high density predictions [equation (\ref{airy}) through (\ref{sp})], we see that in the Wigner crystal, $|\phi(q)|^2$ is more spread out in momentum space, and does not contain any oscillations.





\begin{figure}
\includegraphics[width=\columnwidth]{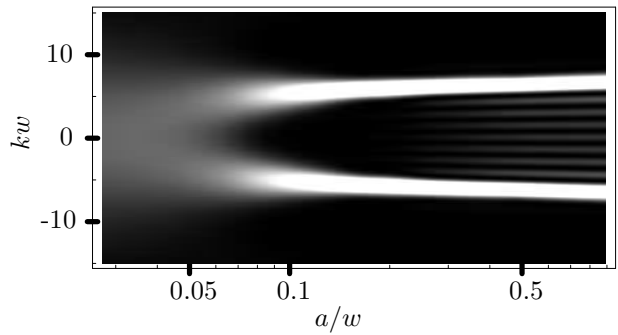}
\caption{Interaction dependence of the
momentum space wavefunction of last single-particle state within the Hartree-Fock approximation.  Eighteen particles are confined to the wire.  Lighter colors correspond to larger values of $|\phi(k)|^2$.  The horizontal axis shows different interaction strengths parameterized by the ratio of the Bohr radius $a$ to the wire length $w$ (note the logarithmic scale). The potential has exponent $\beta=6$, and a cutoff, $d=0.07 w$.
}\label{fig:hf}
\end{figure}




\section{Hartree-Fock study of crossover}\label{hf}
\begin{figure}
\includegraphics[width=\columnwidth]{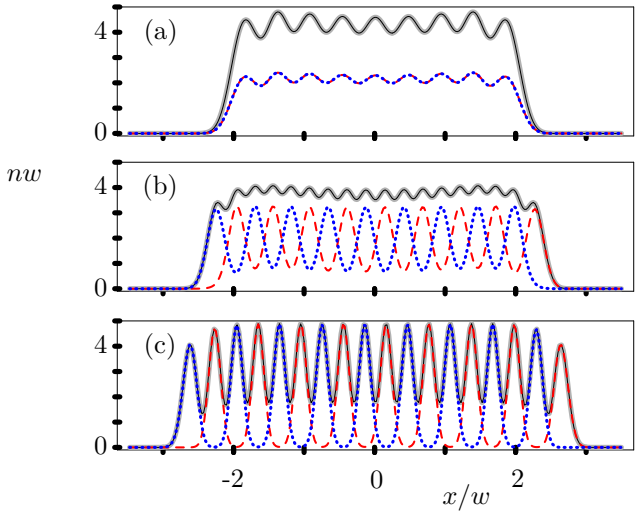}
\caption{(Color Online) Real space electron density $n$.  Parameters coincide with Fig.~\ref{fig:hf}.  Solid, blue dotted, and red dashed lines correspond to total density and the density of up and down spin electrons [In (a), the latter two curves coincide].  Panels (a), (b), (c) correspond to $a/w$=0.5, 0.15, 0.05.
}\label{fig:hf2}
\end{figure}
We  investigate the crossover between the high density electron gas and the low density Wigner crystal
within the Hartree-Fock approximation.  
In the extreme limits, Hartree-Fock reduces to our previous approximations.  It also provides 
 semi-quantitative understanding of the intermediate regime.
  In uniform higher dimensional systems, Hartree-Fock overestimates the stability of the Wigner crystal \cite{trail}, and one therefore expects some systematic errors as compared to an exact many-body calculation.
Unlike bosonization and Luttinger Liquid techniques, Hartree-Fock correctly accounts for band curvature.  Band curvature plays a significant role at low densities.  Appendix~\ref{two} compares Hartree-Fock and exact results for two particles.

Hartree-Fock is most simply thought of as a variational method, where one searches for the Slater determinant which minimizes the energy \cite{ashcroft}.  This procedure is highly nontrivial as the energy landscape in the space of Slater determinants is quite complex with many local minima.  We start by considering the limit of vanishing interactions $(a/w\to\infty)$, where we find the exact solution by discretizing space and solving the single-particle Schrodinger equation.   
With fixed numbers of up and down spin electrons ($n_\uparrow$ and $n_\downarrow$), we then gradually decrease $w/a$.  For each value of $w/a$ we iteratively solve the Hartree-Fock equations in discretized space.  Although there is no guarantee that the states we find in this manner are absolute energy minima,  they consistently have lower energies than all other states that we have found by iterating the Hartree-Fock equations from different starting points.



An advantage of the Hartree-Fock approximation is that since it involves an independent electron approximation, one has direct access to single-particle observables, such as the wavefunction of the last bound state.  Figure~\ref{fig:hf} shows the momentum-space wavefunction of the last bound state for a system of eighteen particles (nine in each spin state).  For large $a/w$, this last single-particle state is delocalized in real space, resulting in a series of momentum-space peaks which are well described by equation (\ref{airy}).  For small $a/w$, this last state is localized, resulting in a broad spread of momenta.  Illustrative real-space density profiles are shown in figures~\ref{fig:hf2} a,b,c.  In (a) we see that even for arbitrarily weak interactions the real space density is corrugated.  The corrugations are caused by the free electron response to an inhomogeneous potential and are analogous to the Friedel \cite{friedel} oscillations seen in electron density near an impurity.  The asymptotic behavior of these oscillations has been analytically calculated for harmonic confinement\cite{ejm}.  There are nine peaks corresponding to the nine doubly occupied single particle states. In (b) we see that as we increase the interaction strength each of the density peaks split into two, so there is one peak per particle.  A spin-density wave appears.  In (c) we see a Wigner crystal with an antiferromagnetic spin profile.  The density varies smoothly as one increases the interactions.

\subsection{Fringes}\label{fringe}
\begin{figure}
\includegraphics[width=\columnwidth]{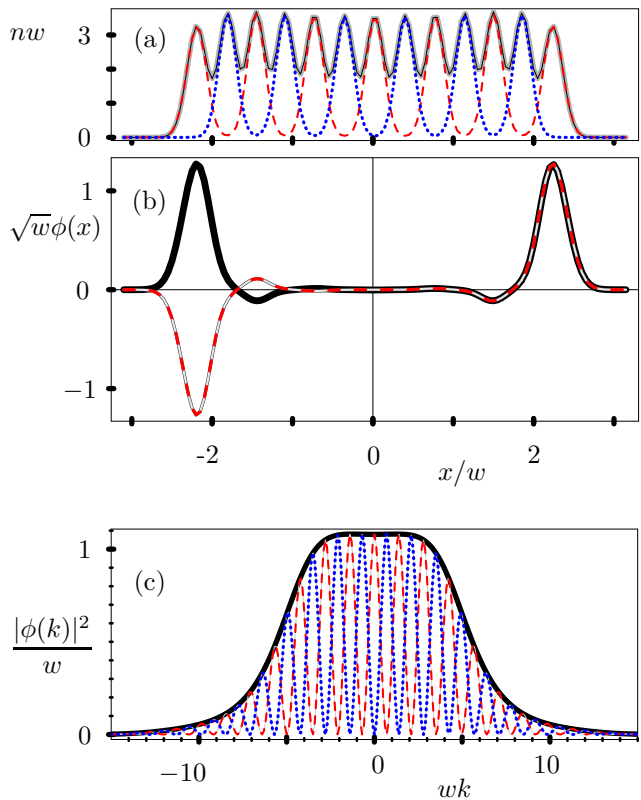}
\caption{(Color Online) (a) Density of 13 electrons with $a/w=0.1$, see caption of fig~\ref{fig:hf2} for key. (b) Wavefunction $\phi_i(x)$ (solid) and $\phi_j(x)$ (dashed) of the two highest highest energy occupied states. (c) Momentum space wavefunctions: red dashed, $|\phi_i(k)|^2$; blue dotted, $|\phi_j(k)|^2$; solid, $(|\phi_i(k)|^2+|\phi_j(k)|^2)$.}\label{symbreak}
\end{figure}

When an odd number of particles form a Wigner crystal the highest occupied single particle state, with wavefunction $\phi_i(r)$, is in a symmetric superposition of two locations, as illustrated in figure~\ref{symbreak}, and has fringes in its momentum space wavefunction, $\phi_i(k)$.  The next highest energy state, $\phi_j(r)$, is in the antisymmetric superposition of these same two locations, and has momentum space fringes which are 90 degrees out of phase with those for $\phi_i(k)$.  Since the splitting between these two states is exponentially small, the tunneling current is proportional to $|\phi_i(k)|^2+|\phi_j(k)|^2$ (see section~\ref{indep}), which, as illustrated in figure~\ref{symbreak}c, contains no fringes. 

Going beyond Hartree-Fock, one generically expects the many-body wavefunction to include amplitudes for each particle to be in a superposition of more than one place.  Thus, in an exact treatment one would always expect that tunneling from any one particular state to any other particular state would result in fringes.  However, as with the Hartree-Fock result, if one incoherently adds the fringes from several symmetry related states then the fringes will be washed out.  In particular, if the temperature or lifetime is greater than the splitting between these states then no fringes will be observable.

\section{Summary}
We have presented simple analytic theories of very high and very low density electrons in an inhomogeneous one dimensional potential.  At high density an electron liquid is formed where the density displays Friedel oscillations from response to the potential [Fig.~\ref{fig:hf2}(a)].  At low densities a crystal-like state is formed where the electrons are well-separated [Fig.~\ref{fig:hf2}(a)]. Using the Hartree-Fock approximation we numerically studied the cross-over between these limits.  Intermediate between the two extremes we found a spin-density-wave state [Fig.~\ref{fig:hf2}(b)].  

In the extreme high density limit our approach is exact.  In the extreme low density limit, our approach will give the exact electronic density.  Our mean-field approach, however, can not describe the quantum spin correlations which occur in the Wigner crystal state.  Instead, it predicts a classical N\`eel order.  One expects that our description of the crossover is qualitatively correct, but that the stability of the crystalline state is overestimated.

Within the independent electron picture we have calculated the tunneling current between electrons in a short and long wire.  In the liquid phase, the zero bias momentum resolved tunneling is dominated by sharp peaks at $\pm k_f$, and there is large overlap between states with different particle number.  As interactions increase, these two peaks broaden and merge. In the crystaline phase, the tunneling displays a large feature centered at $k=0$.  This feature is attributed to electron localization.  Going beyond our mean-field picture, it is clear that in this limit the overlap between states with different particle number becomes extremely small, which leads to coulomb-blockade structures.

For these tunneling predictions we work in linear response, and neglect correlations between the electrons in the two wires.  For quantitative comparison with experiments, one needs to include higher order terms in the tunneling and coulomb induced interwire correlations. 

Although simple mean-field analysis  (sec.~\ref{fringe}) and more general symmetry arguments\cite{halperin} suggest that the large peak at $k=0$ will show a fringe structure, we have demonstrated  that  if either the temperature or the inverse state lifetime $\Gamma$ is larger that the splitting between spin states then this structure is washed out, and may therefore be extremely hard to observe experimentally.  We should emphasize that the Hartree-Fock theory used here is unable to accurately describe these fringes.  From our studies of the two-particle system, we believe that Hartree-Fock does describe the envelope of the peak, though this needs to be confirmed by studying larger systems.


\begin{acknowledgments}
I would like to thank Amir Yacoby for providing experimental details\cite{yacobytalk}, Bertrand Halperin for stimulating correspondence, and Paul McEuen for discussing related experiments on carbon nanotubes.
This work was partially performed at the Kavli Institute for Theoretical Physics.  It was supported in part by the National Science Foundation under Grant No. PHY99-07949.
\end{acknowledgments}

\appendix
\section{Two Particles}\label{two}
The physics of interaction induced electron localization is very simply understood by studying two particles in a parabolic confinement.  The two particle system is particularly appealing because it is equivalent to a problem in one dimensional single particle quantum mechanics and is therefore approachable through elementary means.
The two and three dimensional version of this problem has been extensively studied \cite{3dhocoulomb,2dhocoulomb}.

In section~\ref{exact} we show this exact solution.  In section~\ref{hf2} we solve this problem through the Hartree-Fock method used in section~\ref{hf}.  By comparing these two results one quantifies the accuracy of our approximations. 

\subsection{Exact Solution}\label{exact}
In coordinates scaled by the oscillator length $w$, the two particle Schrodinger equation is 
\begin{eqnarray}
\left[-\frac{\partial_{x_1}^2+\partial_{x_2}^2}{2} +\frac{x_1^2+x_2^2}{2}+\frac{ f_d(x_1-x_2)}{a}\right] \psi(x_1,x_2)\\\nonumber=E \psi(x_1,x_2),
\end{eqnarray}
where the energy $E$ is scaled by the oscillator energy $\hbar\omega=\hbar^2/m^* w^2$.  This equation separates into center of mass an relative coordinates $X=(x_1+x_2)/2$, $x=x_1-x_2$ by writing $\psi(x_1,x_2)=\Psi(X)\psi(x)$, which obey
\begin{eqnarray}
\left[
(-\partial_X^2/4+X^2
\right] \Psi(X)&=& E_X \Psi(X)\\
\left[
-\partial_x^2+x^2/4+f_d(x)/a
\right] \psi(x)&=& E_x \psi(x), \label{spse}
\end{eqnarray}
with $E=E_x+E_X$.  Up to normalization, the lowest energy center of mass wavefunction is $\Psi_X(X)=e^{-X^2},$ and $E_X=1/2$.
Our task therefore reduces to solving the single particle Schrodinger equation (\ref{spse}).
The electron spin enters in that for a spin singlet state $\psi(x)$ must be even, while for a spin triplet, $\psi(x)$ must be odd.


\begin{figure}
\includegraphics[width=\columnwidth]{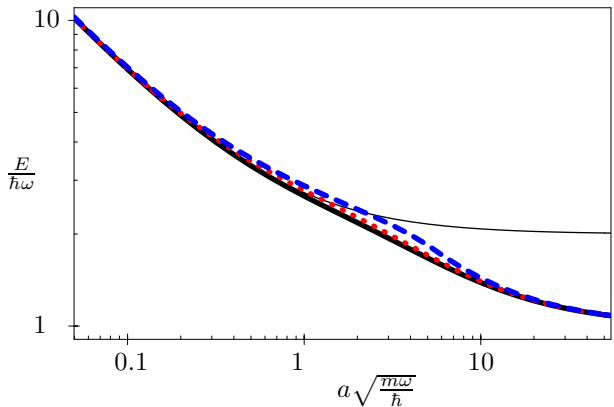}
\caption{(Color Online) Energy of two electrons in a one dimensional harmonic harmonic well. Thick line: exact singlet; thin line: exact triplet; red dotted line: variational singlet; blue dashed line: exact Hartree-Fock. Variational Hartree-Fock is not shown as it is indistinguishable from the blue dashed line. Note logarithmic scale.}\label{exactpsiE}
\end{figure}

\begin{figure}
\includegraphics[width=\columnwidth]{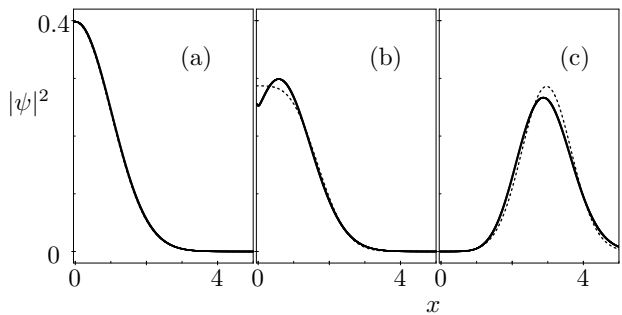}
\caption{Relative wavefunction for two particles in a harmonic well for (a) $a=1000$; (b) $a=10$; (c) $a=0.1$.  All lengths are measured in units of $d=\sqrt{\hbar/m\omega}$.  Solid line shows exact results, while dashed lines show variational approximation.}\label{exactpsiwf}
\end{figure}

Figure~\ref{exactpsiE} shows the energy $E$ of the singlet and triplet states as a function of $a$, while figure $\ref{exactpsiwf}$ shows $\psi(x)$ for several representative values of $a$.  

\subsubsection{Variational Approximation}
Qualitative features of the exact solution are most readily understood through a variational approach.
We take $\psi(x)\propto e^{-(x-t)^2/2s^2}\pm e^{-(x-t)^2/2s^2}$ where $t$ and $s$ are free parameters.  The expectation value of the energy is then
\begin{eqnarray}\label{evare}
\langle E_x\rangle &=& \frac{1}{2 s^2}+\frac{s^2}{8}+\frac{1}{e^{t^2/s^2}\pm 1}
\left(\frac{t^2 e^{t^2/s^2}}{4}\mp\frac{t^2}{s^4}+\frac{g_\pm}{s a}\right)\nonumber\\
g_{\pm}&=&\int_{-\infty}^{\infty} \frac{dx}{2 \sqrt{\pi}}\, f_d(x) e^{-x^2/s^2}\left(
e^{t x/s^2}\pm e^{-t x/s^2}
\right)^2.
\end{eqnarray}
For $d\ll s$ we extract the leading $d$ dependence of $g_\pm$ to arrive at
\begin{eqnarray}
g_- &=& \int_0^\infty \frac{dx\, e^{- x^2}}{x \sqrt{\pi}}\left(e^{\tau x}-e^{-\tau x}\right)^2+{\cal O}(d^2/s^2)\\
g_+ &=& g_-+\tilde g\\
\tilde g &=& \frac{2}{\sqrt{\pi}}\left[ -\gamma-\log \left(d^2/\nu s^2\right)\right]+{\cal O}(d^2/s^2),
\end{eqnarray}
where $\tau=t/s$, $\gamma\approx0.577$ is Euler's gamma and $\nu$ is a constant which depends on the exact form of the regularization.  For regularization $f^{(1)}$, $\nu=e^2$ and for regularization $f^{(2)}$, $\nu=4$.  The energy is only logarithmically sensitive to this constant.

The separation between the two peaks in the wavefunction is given by $t$, and the width of each peak is given by $s$. These play the role that $n^{-1}$ and $\ell$ played in section~\ref{classical}. The lowest energy variational wavefunctions are compared to the exact results in figure~\ref{exactpsiwf}.  By expanding for small $t$, one sees that for any $a<\infty$, the lowest energy variational state always has $t\neq0$.  



\begin{figure}
{
\includegraphics[width=\columnwidth]{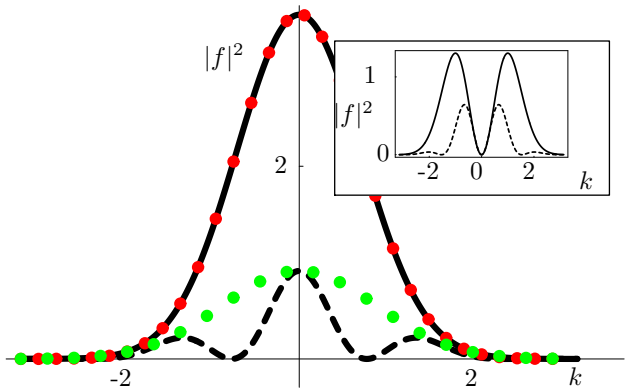}}
\caption{(Color Online)
Tunnelling matrix element $|f(k)|^2$ for two harmonically trapped particles in a singlet state:  exact  $a=100$ (solid), $a=0.05$ (dashed); hartree fock $a=100$ (red dotted), $a=0.05$ (green dotted, scaled by 0.32). All lengths are measured in units of $d=\sqrt{\hbar/m\omega}$.   Inset shows triplet states.
}\label{comparef}
\end{figure}

When appropriately biased, the tunneling current is proportional to $|f(k)|^2$, where $f(k)=\langle 1|\hat \psi(k)|2\rangle$, $k=\sqrt{2m^*{\cal E}_f}-e B d_t/\hbar$, and $|1\rangle$, $|2\rangle$ represent the ground state with $1$ and $2$ particles.  Explicitly,
\begin{equation}
f(k)=\frac{2^{3/4}}{3^{1/2}} \int dx\,\psi(x)e^{-2 i k x/3 - x^2/12-k^2/6}.
\end{equation}
Figure~\ref{comparef} shows this quantity for two illustrative values of $a$.  As with the Hartree-Fock theory in section~\ref{fringe}, one sees fringes from the fact that each electron is in a quantum superposition of two places.  The period (in k-space) of these fringes becomes smaller as interactions increase.  

If one expands the potential in (\ref{spse}) about its minimum, one sees that the size of each peak in the wavefunction becomes  independent of $a$ for sufficiently small $a$.  This is born out by analyzing the variational energy (\ref{evare}) in the limit of large $t$.    Consequently the envelope of the finges in $f(k)$ becomes independent of $a$ for strong interactions.

In a flatter potential ($V\sim x^\beta$ with $\beta>2$, the peaks in $\psi(x)$ will become narrower as $a$ increases, resulting in a broadening of $f(k)$.

\subsection{Hartree Fock}\label{hf2}
We compare our exact results with the Hartree-Fock approximation.  For two particles in the same/different spin states, Hartree-Fock is equivalent to the variational ansatz
\begin{eqnarray}
\psi_{\uparrow\uparrow}(x_1,x_2)&=& (\phi_1(x_1)\phi_2(x_2)-\phi_2(x_1)\phi_1(x_2))\zeta_{1\uparrow}\zeta_{2\uparrow}\\\nonumber
\psi_{\uparrow\downarrow}(x_1,x_2)&=& \left[\phi_1(x_1)\phi_2(x_2)\zeta_{1\uparrow}\zeta_{2\downarrow}-\phi_2(x_1)\phi_1(x_2)\zeta_{2\uparrow}\zeta_{1\downarrow}\right],
\end{eqnarray}
where $\phi_i(x)$ represent single particle orbitals and $\zeta_{j\uparrow}$ corresponds to having the $j$'th particle in the spin up state.  This ansatz clearly leads to a spin triplet state when the two particles have the same spin, but does not lead to any particular symmetry when the particles have dissimilar spins.

We focus on the case of dissimilar spins.
As might be expected from symmetry, we numerically find that the minimum energy configuration always has $\phi_2(x)=\phi_1(-x)\equiv \phi(x)$ [which when $\phi_1(x)$ is even is equivalent to $\phi_2(x)=\phi_1(x)$].  We then find that the minimal energy Hartree-Fock state obeys the self-consistent equation
\begin{equation}\label{fullhf}
\partial_x^2 \phi(x)=(x^2-\mu)\phi(x) +\frac{2}{\bar a} \phi(x)\int \!\! d\bar x\,|\phi(\bar x)|^2 f_d(x+\bar x),
\end{equation}
which is readily solved numerically.  

\subsubsection{Variational}
As with the exact state, we gain extra insight by a variational solution, taking $\phi(x)\propto e^{-(x-t/2)^2/s^2}$, for which
\begin{equation}\label{hfvar}
\langle E\rangle=\frac{t^2}{4}+\frac{s^2}{4}+\frac{1}{s^2}+\frac{1}{2 s \bar a}e^{-t^2/s^2}[g_++g_-].
\end{equation}
Not surprisingly, the interaction term in the energy is a linear combination of the interaction we previously found for triplet and singlet states.

Unlike when we applied the variational approximation to the exact equations (\ref{evare}), here we find that for large $a$, the separation $t$ is strictly zero.  As $a$ decreases there is a transition where $t$ discontinuously jumps to a nonzero value, corresponding to the formation of a spin density wave.  This discontinuity is also seen in the full numerical solution to (\ref{fullhf}), and for $d=0.01$ occurs near $a=4.7$. 

The variational approximation to the Hartree-Fock energy (\ref{hfvar}) is almost indistinguishable from the exact Hartree-Fock energy (\ref{fullhf}).  Figure~\ref{exactpsiE} compares this approximate result to the exact energy from solving (\ref{spse}) and its variational approximation (\ref{evare}).  As we anticipated, the Hartree-Fock result is nearly exact in the limits of very strong and very weak interaction.  For intermediate coupling however the Hartree-Fock energy deviates from the exact result.

Within Hartree-Fock, the tunneling current is proportional to $|\phi_k|^2$, where $\phi_k=\int dx e^{-i k x} \phi(x)$.
Figure~\ref{comparef} compares $|\phi_k|^2$ to the exact result for $|f(k)|^2$.  Since the Hartree-Fock result does not capture the orthogonality catastrophe physics, the two predictions differ by a significant scale factor [0.32 at $a=0.05$].  Furthermore, the exact result contains fringes which are not seen in Hartree-Fock.  These fringes occur because the exact wavefunction places each electron in a superposition of two positions.  Once rescaled, the Hartree-Fock prediction gives the envelope of the exact result.

We expect these observations to be generic.  Even in the system with more particles, the Hartree-Fock prediction should match the scaled envelope of the exact result.  If finite temperature/lifetime effects wash out the fringes then the Hartree Fock should correctly predict the dependance of the tunelling current on momentum (magnetic field).

\section{Multiple-well potentials}\label{doublewell}
\begin{figure}
\includegraphics[width=\columnwidth]{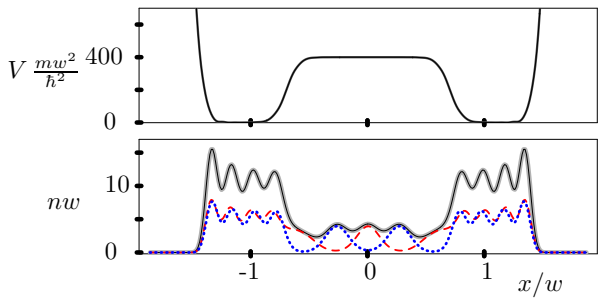}
\caption{(Color Online) Top: Double well potential along wire generated by gates. Bottom: Hartree-Fock electron density in this potential [see fig.~\ref{fig:hf2} for key].  As illustrated, the characteristic length scale of the potential is $w$. The interactions are characterized by $a=0.1 w$ and  $d=0.07w$, yielding a low density crystal-like central region surrounded by a high density  liquid-like region.  }\label{multiwell}
\end{figure}

The power law potentials used in the body of this paper capture the slow variation of electrostatic fields felt by electrons in typical experiments.  Some features of these power-law potentials are overly-simplistic.  Most importantly, due to experimental constraints on the placement of gate electrodes, a multiple-well structure may be formed along the wire, as illustrated in figure~\ref{multiwell}.  Qualitative understanding of the electronic state is found by 
noting that the local chemical potential is small in the central region but large near the edges, resulting in a low density crystaline state surrounded by a high density liquid state.  A Hartree-Fock calculation, illustrated in figure~\ref{multiwell}, verifies this intuitive picture.

The localized states which are formed {\em above} the barrier are a purely many-body effect.  Similar localized state are believed to occur above quantum point contacts\cite{above} and may give rise to the so called 0.7 anomaly\cite{anom}.

\begin{thebibliography}{99}
\bibitem{mufoot} As the chemical potential difference between the wires, $\delta V$ depends not only on the applied potential, but also on the number of electrons in the short wire.
\bibitem{it}
For example, the first correction involves setting $\bar n(\bar r^\prime)= \bar n_0(\bar r^\prime)$ on the right hand side of (\ref{tf2}).
Using regularization $f_2$  this correction is
$({4\pi^2}/{\sigma^2}) (\bar n^2(\bar r)-x \bar n_0)\approx
-(\sigma/{\pi a}) (A+B \bar r^2+{\cal O}(\bar r^4)),$
where $A=
1-{4}/{\beta}+({2}/{\beta})\log({4}/{d_\perp^\beta})$, and
$B=  [n \sqrt{\pi}\Gamma(1-2/\beta)]/[
(n-4)\Gamma(1/2-2/\beta)]$.
\end{thebibliography}
\end{document}